\documentclass[namedreferences,hyperref,optionalrh]{spr-sola}
\usepackage[T1]{fontenc}
\usepackage[utf8]{inputenc}
\usepackage{lmodern}
\usepackage{graphicx,amsmath,amssymb,booktabs}
\usepackage{microtype}
\usepackage{placeins}
\newcommand{\kms}{\,\mathrm{km\,s^{-1}}}
\newcommand{\mk}{\,\mathrm{MK}}
\newcommand{\gauss}{\,\mathrm{G}}
\newcommand{\fast}{\textsc{fast}}
\newcommand{\slow}{\textsc{slow}}
\newcommand{\alfven}{Alfv\'en}
\newcommand{\measured}{\texttt{MEASURED}}
\newcommand{\sourced}{\texttt{SOURCE\_DERIVED}}
\newcommand{\assumed}{\texttt{ASSUMED}}
\newcommand{\modelled}{\texttt{MODEL\_DERIVED}}
\hypersetup{pdftitle={Riemann Map Operator for Solar Front Diagnostics},pdfauthor={Olena Podladchikova},colorlinks=true,citecolor=blue,urlcolor=blue,linkcolor=blue}

\makeatletter
\newenvironment{rmosidefigure}[1][tbp]
  {\@float{figure}[#1]\let\@makecaption\rmo@sidecaption}
  {\end@float}
\long\def\rmo@sidecaption#1#2{%
  \par\figcapfont\raggedright #1\hskip\tabcapspace #2\par}
\def\@doi{}
\def\@copyright{}
\makeatother
\begin{document}
\begin{frontmatter}
\title{Riemann Map Operator for Solar Front Diagnostics}

\author{\inits{O.}\fnm{Olena}~\snm{Podladchikova}}
\institute{National Technical University of Ukraine, Igor Sikorsky Kyiv Polytechnic Institute,\\
37 Beresteiskyi Prospect, Kyiv 03056, Ukraine}
\runningauthor{O. Podladchikova}
\runningtitle{Riemann Map Operator}
\begin{abstract}
Identifying a shock from a moving solar brightness front requires relating the observed emission to changes in the plasma state. The Riemann framework describes shocks, rarefactions and other waves as parts of a pattern connecting different states. We present the Riemann Map Operator (RMO), an inverse diagnostic framework implemented as a web application to help observers identify magnetohydrodynamic (MHD) shock families from incomplete measurements. RMO tests candidate local connections against conservation laws, entropy requirements and characteristic conditions, while retaining observational uncertainties and model assumptions. We demonstrate the approach with five solar examples and a complementary in-situ comparison. An event combining extreme-ultraviolet (EUV) imaging and spectroscopy admits both fast and slow shock connections within the tested conditions. For an EUV--radio event, we derive an upstream-flow constraint that would exclude ordinary slow shocks within the adopted model, provided both diagnostics sample the same front. A flare-loop benchmark reproduces the published spectral profiles and recovers slow-shock ordering in the isothermal limit. The Solar Orbiter comparison recovers an Alfv\'enic relation between measured velocity and magnetic-field changes. Together, the examples show how RMO connects observations to admissible MHD interpretations and identifies the additional measurements needed to distinguish them.
\end{abstract}
\begin{keyword}
\kwd{Magnetohydrodynamics}
\kwd{Waves, Magnetohydrodynamic}
\kwd{Corona, Dynamics}
\kwd{Spectral Line, Diagnostics}
\end{keyword}
\end{frontmatter}

\begin{figure}[!t]
\centering\includegraphics[width=\textwidth]{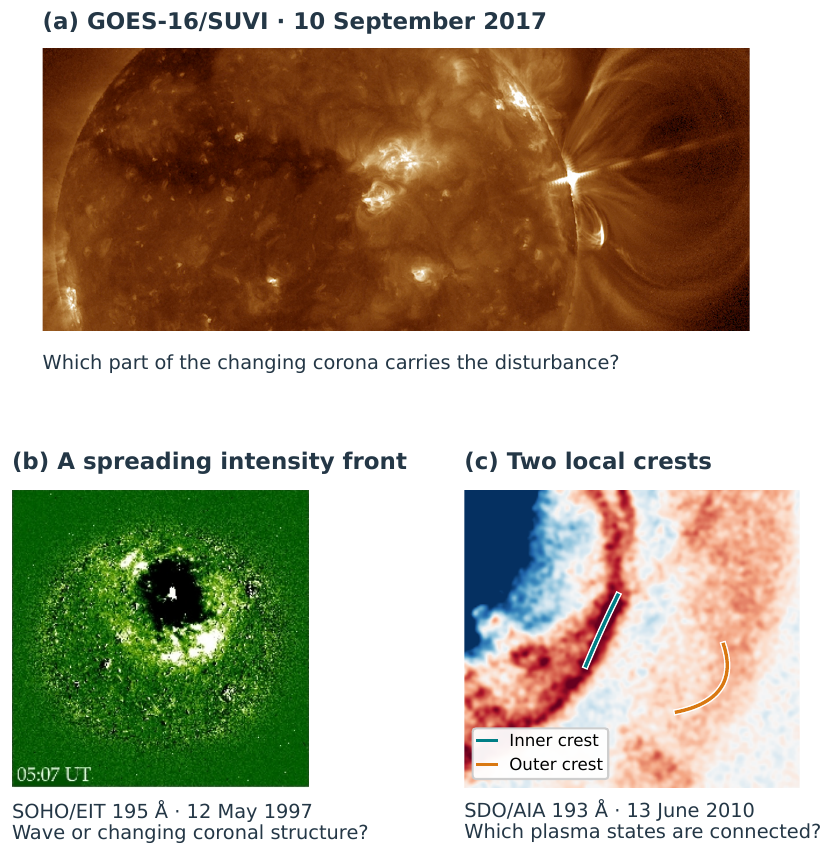}
\caption{What the observer sees. (a) The 10 September 2017 eruption in an official GOES-16/SUVI preview, credited to Dan Seaton, NCEI and CIRES; see \citet{Seaton2018} for the observations and processing. (b) The 05:07 UT frame of the SOHO/EIT 195~\AA{} difference-image sequence on 12 May 1997 \citep{Thompson1998}, from the official SOHO gallery; courtesy of SOHO (ESA and NASA). (c) An AIA 193~\AA{} base-difference image at 05:39:20.05 UTC on 13 June 2010, showing the inner and outer brightness ridges analysed in this study. Blue and orange curves mark their fitted positions, respectively. Panels (a,b) provide context and are not used for quantitative RMO measurements. Panel (c) leads to the local EUV and radio test in Figure~\ref{fig:e11_observation}.}\label{fig:solar}
\end{figure}

\section{Introduction}\label{sec:intro}
A bright, curved front appears in an extreme-ultraviolet (EUV) image of the Sun. It spreads across the disc, rises above an eruption, or separates into several brightness ridges. What physical disturbance does the observer see? Nearly three decades after the early SOHO/EIT observations \citep{Delaboudiniere1995,Thompson1998}, modern imagers resolve far more of this evolution, from the extended corona seen by GOES/SUVI \citep{Seaton2018} to small-scale quiet-Sun brightenings resolved by Solar Orbiter/EUI \citep{Rochus2020,Berghmans2021}. Figure~\ref{fig:solar} starts from what is visible before assigning a physical name.

A tracked ridge gives projected motion and a brightness change gives an emission response. Neither alone establishes whether the feature is a wave, a shock, an expanding material boundary or a changing line-of-sight superposition. EUV disturbances can contain both propagating waves and restructuring associated with a coronal mass ejection (CME) \citep{Zhukov2004,PatsourakosVourlidas2012}, and the long-standing debate over EIT waves already showed that speed and pulse shape alone do not uniquely determine an MHD mode \citep{WillsDavey2007,Nakariakov2005,LiuOfman2014}. By considering speed together with its evolution, \citet{WarmuthMann2011} found evidence for distinct classes of coronal EUV disturbances with different physical interpretations. Passband, cadence, feature selection and projection all shape the measured trajectory, as discussed in the companion perspective \citep{PodladchikovaEIT2026}. Here we ask which physical connections between plasma states those measurements can support.

Images, spectra, radio emission and in situ data constrain different pieces of that question. Spectra provide line-of-sight velocities and emission-weighted plasma properties. Radio diagnostics may constrain density, compression and source location, subject to an emission model and propagation effects \citep{Chrysaphi2018}. Nonthermal radio and hard X-ray emission also trace energetic electrons \citep{PickVilmer2008,Krucker2008}, while in situ instruments can measure plasma and magnetic vectors directly. We can combine these constraints in a local MHD test only after linking them to the front being studied and the plasma on both sides of it.

The calculus of \citet{Newton1736} and \citet{Leibniz1684} provided a mathematical language for describing how physical quantities change. In wave physics, a familiar starting point is to consider small perturbations about equilibrium and retain terms linear in their amplitude. This approach has been highly successful: it lets us study one wave mode at a time and relate its behaviour to the properties of the medium. The small-amplitude assumption is an additional approximation. The full nonlinear equations also describe finite changes, with the medium itself moving, compressing and expanding as the disturbance evolves.

\citet{Riemann1860} studied finite-amplitude sound, extending the treatment beyond the small pressure changes commonly considered at the time. He followed density, pressure and velocity together, and described how a compression can steepen because different parts of the disturbance propagate at different speeds. A shock can thus form as the gas evolves.

\begin{figure}[H]
\centering\includegraphics[width=\textwidth]{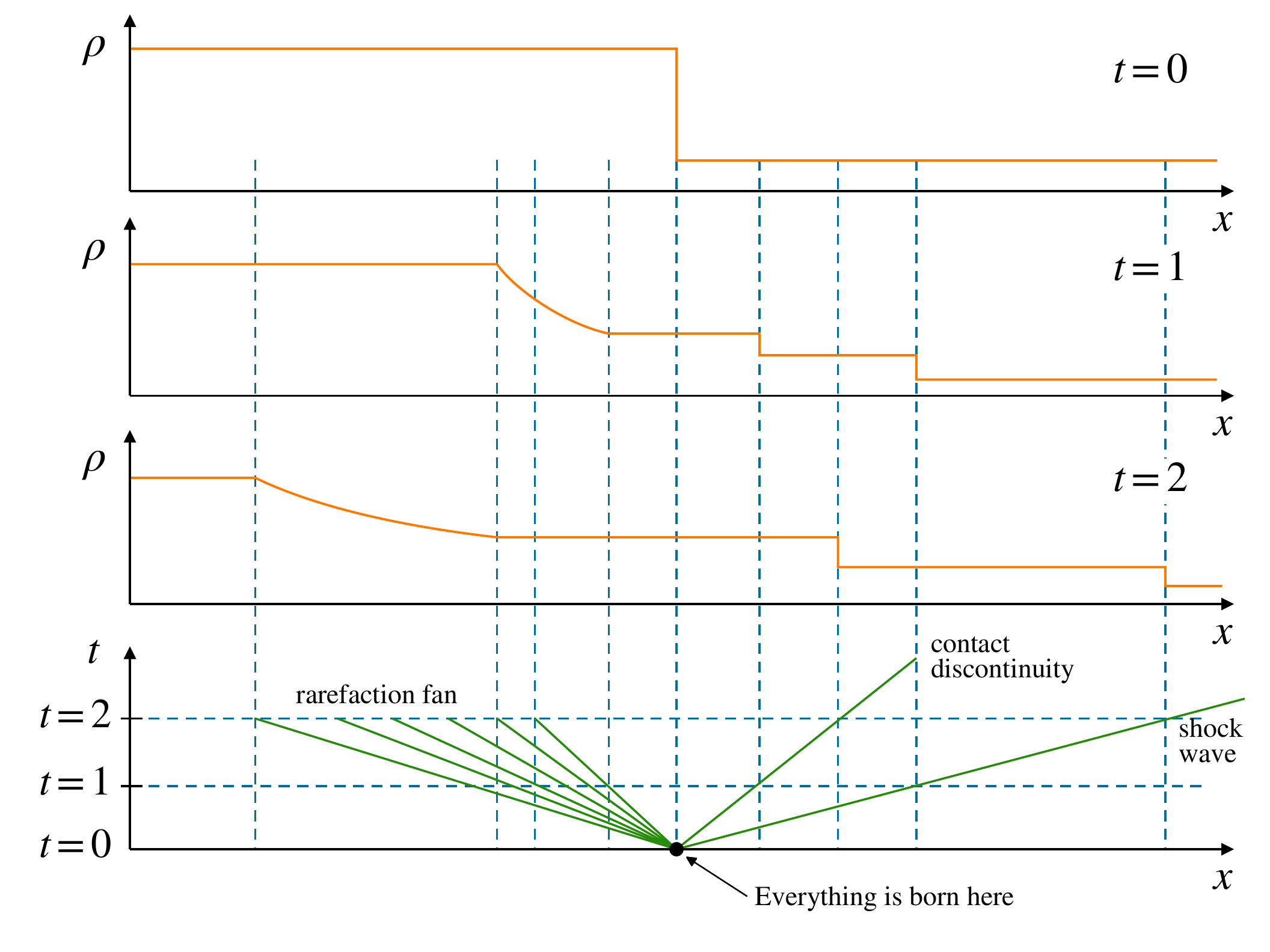}
\caption{A gas-dynamic Riemann fan and illustrative density profiles. The orange curves show an initial density jump and its subsequent evolution: a smooth rarefaction, two intermediate states separated by a contact, and a shock. The green lines below show their positions in time; the blue dashed guides link the profiles to the fan. Density alone does not specify the initial gas states: pressure, velocity and an equation of state are also needed. The times and profiles are schematic, with no calibrated units or calculated solar solution.}\label{fig:hd_fan}
\end{figure}

The modern Riemann problem makes this picture concrete. Consider a long tube containing uniform gas in one state on the left and another on the right, initially separated by a partition. Removing the partition lets the gas evolve. One possible pattern contains an expanding rarefaction, a contact moving with the gas and a shock, with new states between them \citep{Toro2009}. The upper profiles in Figure~\ref{fig:hd_fan} show how the density changes across the tube at successive times. Below them, the positions of the waves and boundaries spread out in time to form a \emph{Riemann fan}. The dashed guides connect this evolving pattern to the profiles above.

Adding a magnetic field introduces magnetic pressure and tension, allowing additional ways for disturbances to propagate. The resulting MHD fan can contain more wave families and intermediate states; a single shock may be one element of this wider pattern \citep{Torrilhon2003}. As in gas dynamics, the initial states and the governing equations determine the possible patterns.

Modern images show fronts and changes in the surrounding corona developing during the same eruption \citep{Seaton2018}. For such scenes, the Riemann picture offers an intuitive starting point: consider the changing medium as a whole, then examine the transitions within it. This view helps place individual wave families in the evolution that the observer sees.

RMO applies this picture to the inverse observational problem. From associated measurements around a selected front, it tests which local MHD connections remain admissible under stated assumptions and which additional measurements would distinguish them. Sections~\ref{sec:concept} and \ref{sec:method} develop the physical description and method; Section~\ref{sec:events} applies it to five solar examples and a complementary in situ comparison.

\FloatBarrier
\section{From Riemann's Discontinuities to Solar Fronts}\label{sec:concept}
To interpret a solar front, we need both the local wave speeds and the conditions linking the plasma on its two sides. CME expansion or a flare-associated pressure pulse can drive the compression that leads to shock formation \citep{VrsnakCliver2008}. We introduce these local properties and then place them within the wider Riemann picture.

\subsection{Wave Speeds Depend on the Plasma State}
In ideal MHD, \emph{characteristics} are the paths or surfaces along which small disturbances propagate. They belong to different wave families. Away from degenerate limits, fast and slow magnetosonic waves involve compression, whereas \alfven{} waves involve transverse magnetic and velocity changes \citep{Benz2002,Aschwanden2005}. Their speeds depend on the local plasma state and the propagation direction relative to the magnetic field; coronal seismology uses this dependence to infer plasma properties \citep{Nakariakov2005}. In a structured plasma, waves can combine properties of these local families, as illustrated by surface \alfven{} waves in magnetic flux tubes \citep{Goossens2012}. We therefore distinguish local characteristic speeds from the modes of an extended solar structure.

A simple wave is a special smooth nonlinear solution that evolves along one such family \citep{Mann1995}, while a general nonlinear disturbance may contain several families and finite discontinuities. The word ``slow'' therefore does not mean simply the slower of two image tracks.

For a uniform state with density $\rho$, pressure $p$ and magnetic field $\mathbf B$, the ideal adiabatic fast and slow characteristic speeds along a unit vector $\mathbf n$ normal to the front are
\begin{equation}
 c_{f,s}^{2}=\frac{1}{2}\left[a^{2}+v_A^{2}\pm
 \sqrt{(a^{2}+v_A^{2})^{2}-4a^{2}v_{A,n}^{2}}\right],
 \label{eq:chars}
\end{equation}
where $a^{2}=\gamma p/\rho$, $v_A^{2}=|\mathbf B|^{2}/(\mu_0\rho)$ and $v_{A,n}^{2}=(\mathbf B\cdot\mathbf n)^{2}/(\mu_0\rho)$. Here $\gamma$ is the adiabatic index and $\mu_0$ is the vacuum magnetic permeability; the plus and minus signs give $c_f$ and $c_s$, respectively. These are plasma-frame characteristic speeds, not projected image speeds; in particular, a linear slow-wave speed is not a universal upper bound on a finite slow shock.

\subsection{A Discontinuity Connects Two States}
A discontinuity is an idealised abrupt change between two plasma states. An ordinary adiabatic shock\footnote{Here \emph{adiabatic} means that heat gained or lost through conduction and radiation can be neglected in the energy balance across the front. The plasma is still heated by compression and irreversible processes within the shock.} compresses the plasma, carries mass through the interface and produces entropy. The plasma enters from the upstream side and leaves on the downstream side. A contact has no mass crossing, while an \alfven{}ic rotational discontinuity rotates the magnetic field and changes the transverse velocity without a density or pressure jump.

The states on the two sides of a shock cannot be chosen independently. Conservation of mass, momentum and total energy, together with magnetic induction and continuity of the normal magnetic field, gives the Rankine--Hugoniot jump conditions. Entropy increase and the ordering of flow and characteristic speeds provide further conditions for an admissible shock. An \emph{evolutionary} check asks whether the discontinuity can respond consistently to small perturbations through the available outgoing waves and jump conditions \citep{Falle2001}.

Here \emph{ordinary} fast and slow shocks denote the nondegenerate compressive branches with regular characteristic transitions. Parallel shocks have the magnetic field along the shock normal on both sides; in switch-on/off shocks, its tangential component appears or vanishes across the front. These cases and coincident-characteristic limits, where characteristic speeds become equal, require separate treatment.

\begin{figure}[t]
\centering\includegraphics[width=\textwidth]{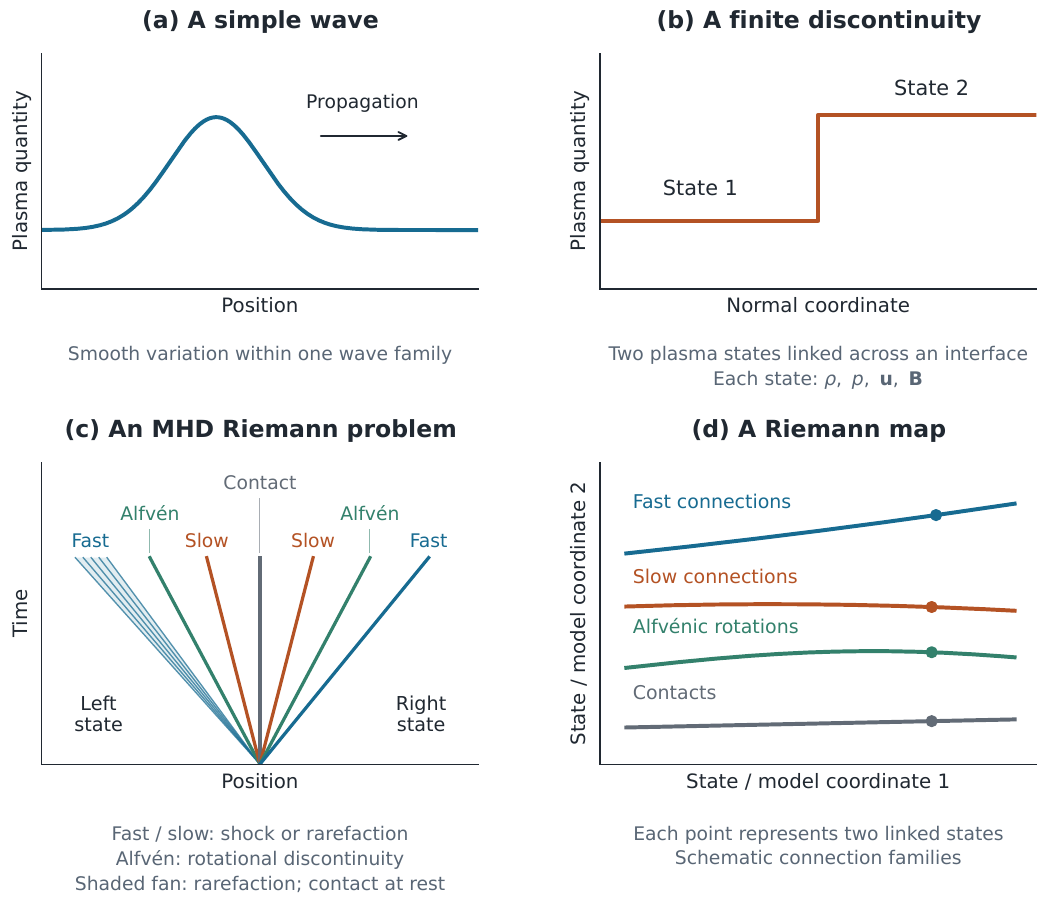}
\caption{What the equations distinguish. (a) A simple wave is a smooth nonlinear solution associated with one characteristic family. (b) A discontinuity links two plasma states through jump and admissibility conditions; only one plasma quantity is shown. (c) A regular ideal-MHD Riemann pattern connects two initial states through waves, discontinuities and intermediate states. Fast and slow waves may be shocks or rarefactions; the \alfven{}ic branches are rotational discontinuities. The shaded fan illustrates a rarefaction. The frame moves with the contact, which therefore appears stationary. (d) A Riemann map represents admissible connections in state or parameter space; each point denotes a connection between two plasma states. Coordinates may represent, for example, density, flow speed or field orientation. All panels are conceptual. The shapes and relative positions of the curves in (d) are schematic, not calculated solar results.}\label{fig:concept}
\end{figure}

\subsection{The Riemann Problem and Its Map}
The Riemann problem follows the evolution from two initially uniform states, as illustrated for a gas in Figure~\ref{fig:hd_fan}. In ideal MHD a regular one-dimensional solution can contain fast and slow shocks or rarefactions, \alfven{}ic rotational discontinuities and a contact, with intermediate plasma states between them. Each shock connects neighbouring states within this evolving pattern. A visible front, once associated with a physical transition, may trace one such connection.

The same initial states can admit more than one MHD wave pattern, depending on the admissibility conditions \citep{Torrilhon2003}. The exact solver of \citet{TakahashiYamada2014} explores these alternative solutions.

A \emph{rarefaction} is a smooth expansion in which characteristics spread apart, whereas an ordinary adiabatic shock is a compressive, entropy-producing transition. A shock and a rarefaction can occur in the same Riemann solution, but their presence and ordering depend on the states being connected. A bright solar front alone therefore does not imply a trailing rarefaction. Interpreting a dark region as a rarefaction requires evidence of the corresponding plasma expansion. Density depletion, plasma evacuation, temperature changes and line-of-sight effects can all reduce EUV intensity \citep{Zhukov2004,Vanninathan2018}. Early Yohkoh observations linked soft X-ray dimming to the loss of coronal plasma during an eruption \citep{Hudson1996}. The plasma response, rather than the sign of the brightness change alone, determines the physical interpretation.

A \emph{Riemann map}, as used here, represents admissible connections or solution families in a space of physical states and uncertain parameters. Its coordinates may include geometry, upstream motion or thermodynamic quantities. RMO uses this representation to organise the local connections allowed by incomplete observations. Here we test individual jumps between neighbouring states; we do not reconstruct an eruption's complete wave pattern.

A \emph{shock polar} addresses a narrower geometrical problem: for a specified upstream state and shock relation it traces compatible downstream states, often through flow speed and deflection \citep{Urashima1966}. A shock polar and a Riemann map use the same jump and admissibility physics, but a map can also vary the uncertain upstream conditions.

\subsection{Solar Structure and Forward Models}\label{sec:forward}
Applying these connections to the Sun requires information about the surrounding coronal structure. Eclipse images reveal that structure close to the Sun, while coronagraphs follow streamers, ejections and faint CME fronts over time. SOHO/LASCO and Solar Orbiter/Metis observe the corona, and the coronagraphs and heliospheric imagers in STEREO/SECCHI extend the view into the heliosphere \citep{Mikic2018,Brueckner1995,Antonucci2020,Howard2008,Eyles2009}. Excess white-light brightness can constrain compression when the scattering geometry and line-of-sight depth are specified \citep{OntiverosVourlidas2009}, but morphology and kinematics alone do not determine a local MHD family.

Models help explain how disturbances arise in this structured corona. Magnetic complexity is statistically related to eruptive activity \citep{Georgoulis2008}, while magnetic topology describes how field lines connect different regions and helps identify possible reconnection sites \citep{Demoulin2006}. The separatrix-web (S-Web) model links open and closed magnetic regions to slow-wind release \citep{Antiochos2011}. \citet{Reville2020} used PLUTO with the HLLD Riemann solver
\citep{MiyoshiKusano2005} to model how tearing can generate
recurrent density structures. Wave models connect plasma
dynamics with coronal heating and solar-wind acceleration
\citep{Ofman2010,VanDoorsselaere2020}. Driven MHD turbulence can concentrate magnetic dissipation in intermittent current sheets \citep{Einaudi1996}. Models also examine how turbulence, reconnection and shocks can act together in heating and particle acceleration \citep{VlahosIsliker2019}. \citet{AnastasiadisVlahos1994} modelled particle acceleration through repeated encounters with shocks in an evolving active region.

Predictive Science's MAS model evolves the global coronal field and plasma and can generate synthetic eclipse, white-light and EUV observables \citep{Mikic2018}. MAS has also been used to simulate global EUV waves and to test how reliably their speeds and plasma properties can be recovered from synthetic observations \citep{Downs2021}. ENLIL and EUHFORIA propagate the solar wind and disturbances through the heliosphere from specified boundary conditions \citep{Odstrcil2003,Pomoell2018}.

These are forward problems: specified states and boundary conditions determine an evolution and its predicted observables. A familiar example from gas dynamics is the calculation of shock waves around a supersonic aircraft such as Concorde. Solar MHD simulations also account for the magnetic field. In both gas dynamics and MHD, Riemann solvers can form part of the numerical method: they use the states in neighbouring computational cells to calculate local fluxes across their shared boundary \citep{Toro2009,TakahashiYamada2014}.

RMO asks the complementary inverse question: when observations supply only part of the local state, which MHD connections remain possible, and what would distinguish them? Forward models can contribute geometry, fields or plasma states, with their assumptions retained explicitly. Keeping observations, model assumptions and physical inference distinct is also central to broader coronal-heating diagnostics \citep{Klimchuk2006}.

\section{From a Visible Front to an MHD State Connection}\label{sec:method}
We start from a moving brightness front and ask whether it can be interpreted as a shock. Ideally, we would measure density $\rho$, pressure $p$, velocity $\mathbf u$ and magnetic field $\mathbf B$ on both sides of the same front patch. These measurements describe two neighbouring plasma states, which may lie within a wider wave pattern. The front's orientation and motion then allow us to test the local transition between them.

In a one-dimensional description, these two states are often labelled \emph{left} and \emph{right}. They lie on opposite sides of the chosen interface; the labels do not mean left and right in the solar image. For a shock, we identify the upstream side by the plasma entering the front and the downstream side by the plasma leaving it, both relative to the moving front.

To prepare a local test, an observer selects the front patch and records its motion and geometry. The spectra, images and magnetic information associated with each side constrain the two states. Their locations, times and uncertainties must be kept with them. Missing properties enter as stated assumptions or ranges; linked quantities, such as temperature and an inferred density, are varied together. RMO tests the resulting connections and identifies the families that remain possible. The RMO web application allows readers to explore this diagnostic workflow interactively at
\url{https://rmo-solar.org/}.

A full reconstruction is not always needed. Within a specified model, an independent bound on plasma flow may exclude a shock family even while some magnetic components remain unknown. If several connections survive, their differences guide the next measurement. Figure~\ref{fig:inference} summarises this sequence; Section~\ref{sec:events} follows it through the solar examples.

\subsection{Association Comes Before State Construction}
Different observations of the same eruption do not necessarily follow the same structure. We therefore check whether a leading intensity crest, an expanding material boundary, a radio source and a Doppler component belong to the same front patch. Likewise, a pre-event reference spectrum does not necessarily represent the upstream plasma of a later discontinuity. In the 1997 event sample of \citet{Klassen2000}, EIT waves and type~II bursts were closely associated, although their inferred speeds were not correlated. Radio imaging can help separate electron-acceleration sites within one eruption \citep{Carley2016}. Combining radio, EUV and white-light observations can also relate these sites to different parts of a CME and its surroundings \citep{SalasMatamoros2016}.

For each constraint, we record the sampled feature, aperture, exposure, reference and geometric convention. The assignment of samples to the upstream and downstream sides must be physically justified. We also specify the plasma composition and equation of state, which links temperature to pressure and density.

If $V_n$ is the front speed along its local unit normal $\mathbf n$ and $u_{1n}=\mathbf u_1\cdot\mathbf n$ is the upstream plasma velocity component in the same direction, the incoming speed relative to a front overtaking the plasma is
\begin{equation}
 w_1=V_n-u_{1n}>0.\label{eq:flow}
\end{equation}
A speed measured along a selected image path does not necessarily equal $V_n$. Even when $V_n$ is known, $u_{1n}$ is needed to obtain $w_1$. Combining motion and emission diagnostics, \citet{DeGroof2004} interpreted moving EIT brightenings as plasma downflows rather than slow waves. In radiative-MHD simulations of spicules, heating fronts produce apparent motions much faster than the plasma flow \citep{DePontieu2017}.

\begin{rmosidefigure}[t]
\centering
\begin{minipage}[c]{0.74\textwidth}
\includegraphics[width=\linewidth]{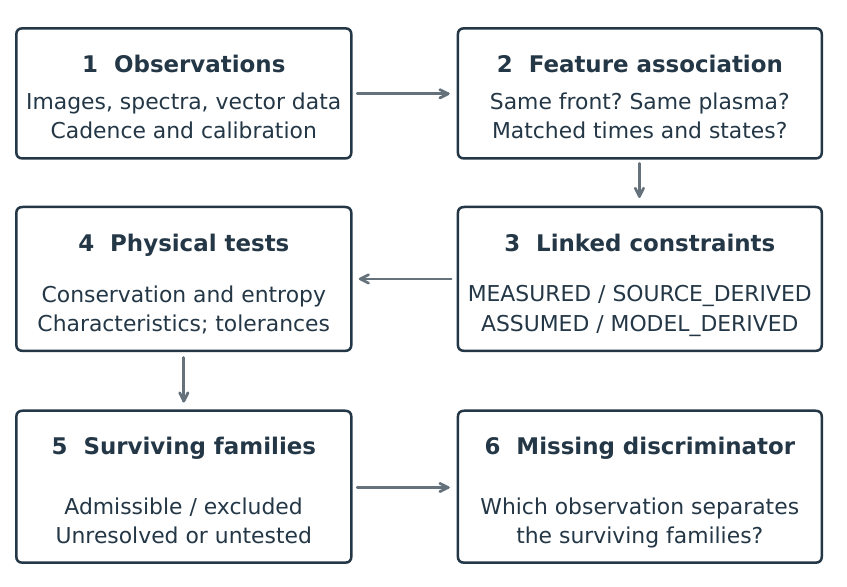}
\end{minipage}\hfill
\begin{minipage}[c]{0.23\textwidth}
\raggedright
\caption{From observations to a physical test. The numbered steps link feature association and the sources of the inputs to physical checks, surviving alternatives and the next useful measurement. Step 3 keeps measurements, quantities derived from observations, assumptions and model results distinct (Section~\ref{sec:inputs}).}\label{fig:inference}
\end{minipage}
\end{rmosidefigure}

\subsection{Sources and Shared Dependencies}\label{sec:inputs}
Four classes describe how each input or result was obtained.

\measured{} denotes detector observations and explicitly identified direct observational diagnostics.

\sourced{} denotes quantities obtained through observational analysis, such as line ratios, registered tracks or compression inferred from split radio-emission bands.

\assumed{} denotes an assumption used to complete the plasma description, such as an equation of state, stationary plasma ahead of the front, a temperature scenario or the assignment of a reference to the upstream state.

\modelled{} denotes quantities produced by a physical model or calculation, including reconstructed magnetic fields, characteristic speeds and conditional exclusion thresholds.

The class is independent of who supplied the quantity. A published simulation remains model-derived, and a measured spectrum remains observational when supplied by another group. Small formal fitting errors do not remove calibration uncertainties or model dependencies.

Inputs derived from the same measurement or assumption must remain linked. For example, a density inferred from an atomic response depends on the assumed temperature. Radio-derived quantities may share assumptions about the emission harmonic and the assignment of spectral bands to plasma states. Analyses using different reference spectra can share the same target exposure. Their joint uncertainty cannot be replaced by independent error ranges without justification. We use a \emph{supported domain} for observationally justified joint constraints and an \emph{assumed domain} for an explicitly declared set of source or model alternatives. These domains specify which input combinations we consider, but not how likely each combination is.

\subsection{Physical Checks and Robustness}

We specify the equations, assumptions and types of connection to be tested between the two plasma states. The physical checks determine which connections are admissible and which MHD families they belong to.

For ordinary adiabatic shocks, the checks include positive density and pressure, compression and all relevant flux balances. We also require entropy increase and independently compare the normal plasma speed in the shock frame with the MHD characteristic speeds on each side. The normal magnetic field must be continuous across a locally planar discontinuity. Appendix~\ref{app:checks} summarises these conditions and the numerical safeguards.

A result is robust over a specified input range if it continues to hold as those inputs change together. Numerical tests establish results for the sampled combinations, whereas an analytical proof can establish a result throughout a continuous domain. Neither gives the probability or observed frequency of a shock family.

For a defined local problem, we distinguish three claims: \emph{checked and admissible}; \emph{checked and excluded}; and \emph{not exhaustively searched}. An admissible example establishes that its family is possible within the tested conditions. Exclusion requires an applicable physical check; failure to find an example is not sufficient. Boundaries within the numerical guard regions remain unresolved.

If the available measurements and stated assumptions do not define a local problem for the two sides of the front, the family test cannot yet be performed.

\FloatBarrier
\section{Solar Demonstrations}\label{sec:events}
During development, RMO recovered the expected classifications for four types of MHD discontinuity in model tests: ordinary fast and slow shocks, \alfven{}ic rotational discontinuities and contacts. We also varied the input parameters within specified uncertainty ranges to test the stability of these identifications.

We now revisit five previously studied solar examples with different observational constraints. Their published observations and models provide the starting point for our analysis. Table~\ref{tab:events} lists the main previous studies, RMO results and information needed next. We retain the identifiers from a larger catalogue of selected solar fronts, so the numbers are not consecutive and do not indicate the order of presentation.

The first two examples provide the main shock comparison: can the measurements distinguish ordinary fast and slow shocks as explanations of a candidate compression? The second also examines contact and rotational discontinuities, together with candidate shocks in which the tangential magnetic field reverses direction.

In the next two examples, the observations do not establish the plasma states needed for a local test, so the MHD family remains unresolved. A flare-loop benchmark then tests a published slow-shock interpretation using spectra and a simulation cut. Finally, a complementary Solar Orbiter comparison examines the \alfven{}ic relation between measured velocity and magnetic-field changes, showing what direct vector measurements add.
\begin{table}[!htbp]
\newcommand{\tc}[1]{\parbox[t]{\linewidth}{\raggedright #1}}
\caption{Solar examples, previous studies and RMO results. Each interpretation depends on the stated geometry, assignment of plasma samples and thermal model.}\label{tab:events}
\fontsize{8.5}{10.5}\selectfont
\setlength{\tabcolsep}{3pt}
\renewcommand{\arraystretch}{1.12}
\begin{tabular}{@{}p{.18\textwidth}p{.16\textwidth}p{.185\textwidth}p{.205\textwidth}p{.20\textwidth}@{}}
\toprule
\tc{Example and studies} & \tc{Observational constraint} & \tc{Result within stated conditions} & \tc{Main limitation} & \tc{Information needed}\\
\midrule
\tc{\textbf{E05}\newline 16 Feb. 2011\newline \citet{Veronig2011}} & \tc{Relative EIS response; AIA brightening} & \tc{Conditional FAST and SLOW} & \tc{Unmeasured local geometry, flow and state assignment} & \tc{Associated geometry/flow or sign-resolved magnetic jump}\\
\addlinespace
\tc{\textbf{E11}\newline 13 June 2010\newline \citet{Ma2011,Kozarev2011}} & \tc{Outer EUV ridge; radio band splitting} & \tc{Conditional FAST; SLOW at other assumed flows} & \tc{No normal-flow bound or local radio/EUV match} & \tc{Local radio match and $u_{1n}<277.5$ km s$^{-1}$ within adopted model}\\
\addlinespace
\tc{\textbf{E02}\newline 27 July 2010\newline \citet{ChenWu2011}} & \tc{Distinct evolving image branches} & \tc{Local family test not yet possible} & \tc{Local states and normal undetermined for either ridge} & \tc{Associated plasma response, states and normal for each ridge}\\
\addlinespace
\tc{\textbf{E08}\newline 13 June 1998\newline \citet{Harra2003};\newline Madjarska et~al.\ \citeyearpar{Madjarska2015}} & \tc{Front crossing; later filament spectrum} & \tc{Filament separated from front; family untested} & \tc{Front spectrum unresolved by available sampling} & \tc{Resolved front spectrum and matched states}\\
\addlinespace
\tc{\textbf{E09}\newline 22 June 2015\newline \citet{Ye2026}} & \tc{Reproducible spectra; published model cut} & \tc{Nine isothermal checks support published slow interpretation} & \tc{No unique observed state pair or full conductive verification} & \tc{Signed vectors, front normal/motion and energy fluxes}\\
\addlinespace
\tc{\textbf{Solar Orbiter}\newline 30 Aug. 2021\newline \citet{Suen2023}} & \tc{Direct vectors; reproduced Wal\'en slope} & \tc{Strong positive \alfven{}ic relation} & \tc{One proton sample in boundary core} & \tc{Resolved states, local normal and pressure treatment}\\
\bottomrule
\end{tabular}
\end{table}
\FloatBarrier

\subsection{16 February 2011 (E05) --- Admissible Fast and Slow Alternatives}\label{sec:e05}
E05 brings together images from the Atmospheric Imaging Assembly (AIA) on the Solar Dynamics Observatory (SDO) and spectra from the EUV Imaging Spectrometer (EIS) on Hinode. AIA shows a small brightening in the area sampled by EIS. The Fe~XIII line shows a redshift equivalent to about $17\kms$ relative to the reference spectra. Can these observations distinguish an ordinary fast shock from an ordinary slow shock?

\begin{figure}[!htbp]
\centering\includegraphics[width=\textwidth]{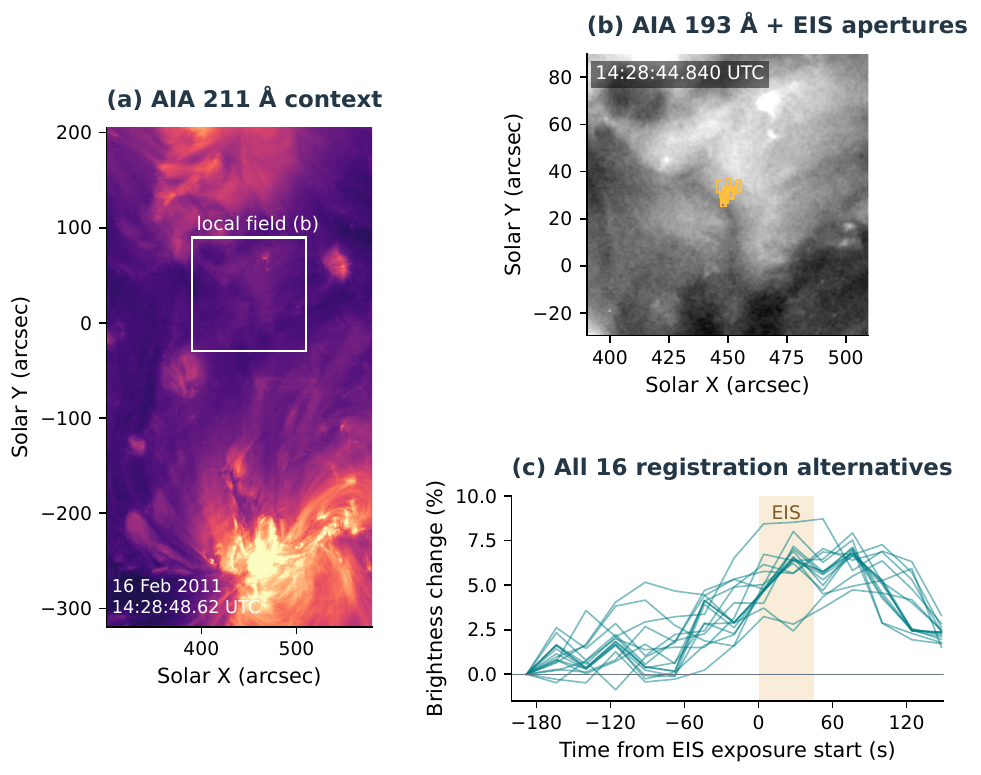}
\caption{E05 observational context, 16 February 2011. (a) AIA 211~\AA{} image; the box locates panel (b). (b) AIA 193~\AA{} during the EIS exposure; the orange rectangles show 16 possible alignments of the same $2\times5$~arcsec aperture. (c) AIA brightness relative to 14:25:32.840 UTC; the shaded interval is the EIS integration. These saved products come from the RMO analysis of the \citet{Veronig2011} event.}\label{fig:e05_observation}
\end{figure}

Figure~\ref{fig:e05_observation} shows where the spectrum was recorded and how the brightness changed there. The result depends on how the images and spectra are aligned and which reference spectrum is chosen. We keep these alternatives linked because they reuse the same observations. The images do not identify a unique local front or its three-dimensional orientation, and the reference spectrum is not securely associated with plasma ahead of that front.

We therefore construct possible states on the two sides using the measured response and explicit assumptions. The density inferred from the spectrum depends on the assumed temperature, so these quantities are varied together. The published mean speeds provide alternative estimates of front motion, while the local geometry and upstream flow must also be specified. RMO completes the candidate states and checks whether the transition between them satisfies the adopted shock conditions.

\begin{figure}[t]
\centering\includegraphics[width=\textwidth]{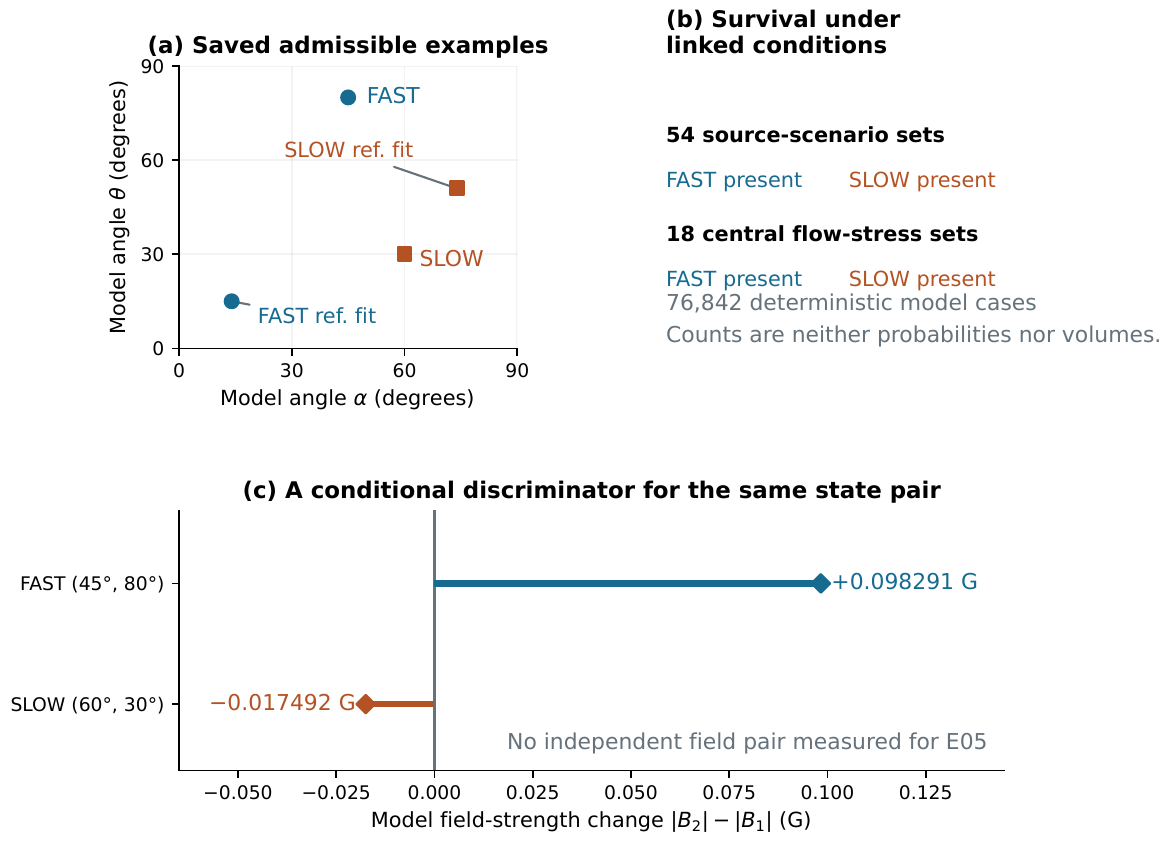}
\caption{Fast and slow alternatives for E05. (a) Admissible model examples at different assumed front-normal tilts $\alpha$ and field-to-normal angles $\theta$. The labelled examples also pass the reference-spectrum test. Blue denotes fast and orange slow; the points do not define continuous boundaries. (b) Both families survive every retained set of input choices. (c) Magnetic-field strength increases in the fast example and decreases in the slow example. These are modelled fields; no corresponding field pair was measured.}\label{fig:e05}
\end{figure}

Within the adopted model, the same observational inputs can be matched by either a fast or a slow shock, depending on the assumed front orientation and the unmeasured plasma properties (Figure~\ref{fig:e05}). Both families have admissible examples in every retained input set. These are alternative interpretations of the same sampled region, with different reconstructed plasma states. Counting the accepted examples does not tell us which interpretation is more likely. Appendix~\ref{app:e05details} gives the input values, sampling and numerical results.

We also check whether the spectra favour one family. A fast example and a slow example both fit the reference spectra nearly as well as a more flexible comparison model. Their front and field directions are assumed. Motion along the front is also assumed to have no line-of-sight component, so it can be present without producing a Doppler shift. These fits neither select a family nor establish that the reference spectrum describes upstream plasma.

What would distinguish the alternatives? If the plasma ahead of the front is stationary, a sufficiently precise measurement of the front's orientation can help: a front-normal tilt around $45^\circ$ excludes the checked slow family, whereas one around $60^\circ$ excludes the checked fast family. These are conditional measurement requirements, not observed angles. Allowing upstream motion restores a slow example at the first tilt and a fast example at the second. Geometry must therefore be combined with information about the plasma flow.

A second route is to measure the magnetic-field strength on both sides of the same front patch. It increases across an ordinary fast shock and decreases across an ordinary slow shock (Section~\ref{sec:discriminators}). Figure~\ref{fig:e05}c illustrates this contrast with two model examples. Such field measurements are not available for E05. Both shock interpretations therefore remain possible, but their differences identify the geometry, flow or magnetic measurements needed to distinguish them.

\FloatBarrier
\subsection{13 June 2010 (E11) --- An Upstream-Flow Test of the Shock Family}\label{sec:e11}
\begin{figure}[!htbp]
\centering\includegraphics[width=\textwidth]{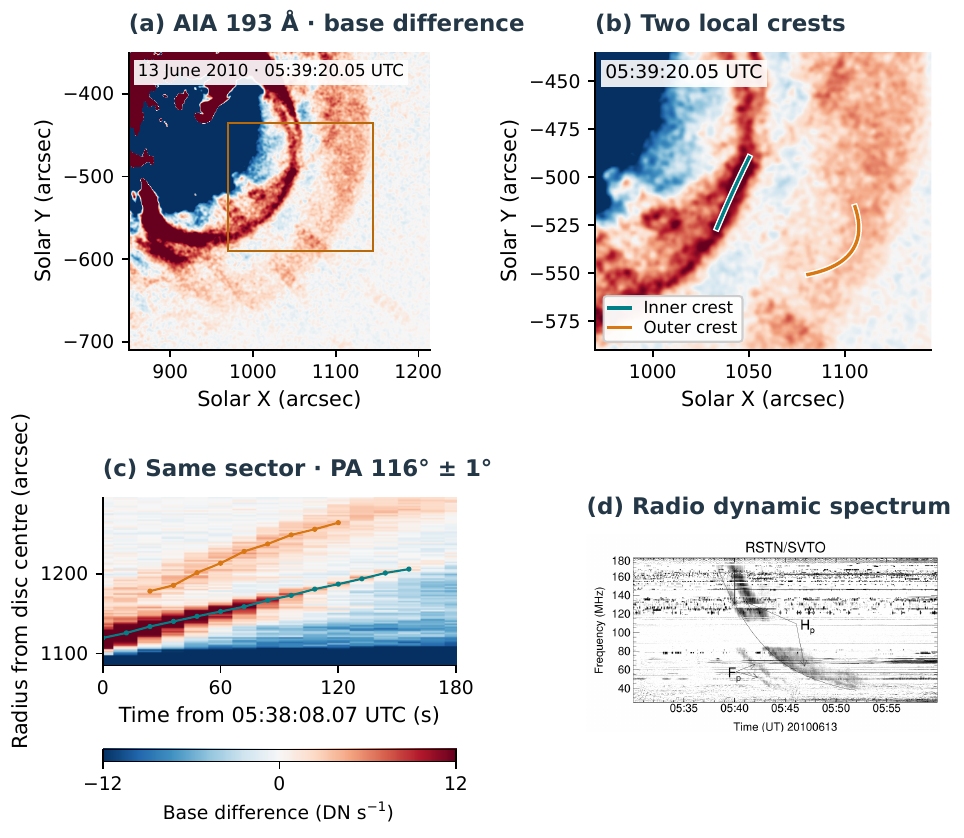}
\caption{E11 observations, 13 June 2010. (a,b) AIA 193~\AA{} base-difference image and local crest fits: blue marks the inner crest and orange the outer crest. (c) Their time--distance tracks in the sector at position angle $116^\circ\pm1^\circ$; the colour scale gives the intensity change in detector counts per second. (d) The RSTN/SVTO radio spectrum reproduced from Figure~7 of \citet{Ma2011}, with the source lane annotations. The spectrum does not localise the radio compression to the selected EUV patch.}\label{fig:e11_observation}
\end{figure}

E11 combines two kinds of measurement: AIA follows an outer EUV brightness ridge, while a split radio-emission band provides a possible estimate of compression. Figure~\ref{fig:e11_observation} shows the separately tracked inner and outer ridges and the radio spectrum. We ask whether the outer ridge and radio measurement can distinguish fast and slow shocks.

To construct the local plasma states, we first adopt a front normal and convert the tracked motion into a normal speed. If the two radio bands come from plasma before and after the same shock, their frequency ratio gives the density ratio. A radio emission model and an assumed temperature then supply density and pressure. The upstream flow and magnetic field remain unknown. We vary these quantities consistently with the local MHD conditions; the precise inputs are listed in Appendix~\ref{app:e11details}.

The resulting fast-shock branch extends over the ordinary, nonparallel field orientations in this model; it does not require a nearly perpendicular field. This is compatible with thermodynamic MHD modelling that identified an outer fast-mode wave separating from the evolving CME structures \citep{Downs2012}. The agreement does not establish that the radio source belongs to our selected EUV patch.

The distinction between fast and slow depends on how rapidly plasma enters the front. The image gives the front's motion, but the incoming speed also depends on motion ahead of it, through Equation~(\ref{eq:flow}). Within the adopted geometry and radio and temperature ranges, an independently measured upper bound
\begin{equation}
 u_{1n}<277.5\kms,\label{eq:threshold}
\end{equation}
would exclude ordinary slow shocks, provided the radio compression belongs to the selected EUV patch. Any measured bound must include its uncertainty. Figure~\ref{fig:e11} shows this threshold and a slow example allowed at a larger assumed upstream flow. Appendix~\ref{app:e11details} gives the derivation and explains how the test changes if the inputs vary together.

\begin{rmosidefigure}[t]
\centering
\begin{minipage}[c]{0.65\textwidth}
\includegraphics[width=\linewidth]{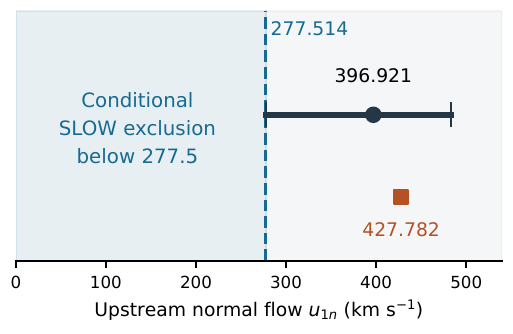}
\end{minipage}\hfill
\begin{minipage}[c]{0.32\textwidth}
\caption{E11 (13 June 2010), with adopted normal speed $V_n=705.535\kms$. The bar spans onset thresholds $277.514$--$483.151\kms$; the circle marks central inputs and the square a checked \slow{} example. A bound $u_{1n}<277.5\kms$ would exclude ordinary \slow{} throughout the adopted domain, provided the radio compression belongs to the selected EUV patch. Neither condition is established.}\label{fig:e11}
\end{minipage}
\end{rmosidefigure}

Neither the flow bound nor the local radio/EUV association has been established. In the adopted geometry the normal lies in the image plane, so a Doppler velocity alone cannot supply the needed flow component. A negligible upstream speed is also an assumption in the standoff analysis of \citet{Gopalswamy2012}; a magnetic field inferred using that assumption cannot independently verify it. Radio frequency drift traces front motion through a density model, rather than the upstream material flow. The test therefore needs an independent flow constraint and a radio source located at the selected patch.

The same local model also tests other connections. A contact or ideal rotational discontinuity cannot provide the imposed compression with plasma crossing the front. The examined candidates with reversed tangential magnetic field fail the ideal-MHD evolutionary test (Appendix~\ref{app:e11details}). These exclusions apply to the adopted states and their assumed radio/EUV association.

Finally, the inner ridge remains a separate feature. A saved slow example can be constructed if the plasma between the ridges is assigned the outer model's downstream state. That shared state is not measured. The two tracks therefore do not establish a fast--slow pair or a complete Riemann fan. For the outer ridge, the useful result is a conditional flow test that observations could complete.

\FloatBarrier
\subsection{27 July 2010 (E02) --- From Two Image Branches to Local Plasma States}\label{sec:e02}
\begin{figure}[!htbp]
\centering\includegraphics[width=0.84\textwidth]{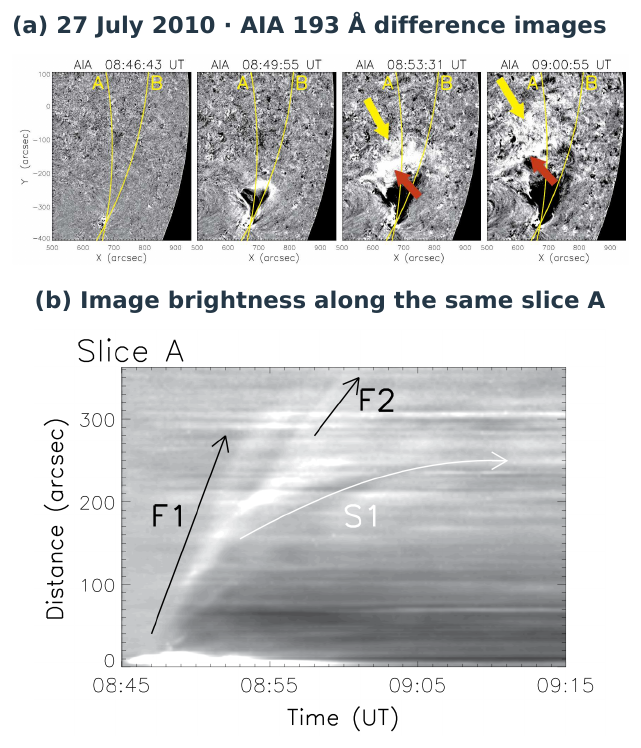}
\caption{E02: seeing the branches before interpreting them. (a) Published AIA 193~\AA{} difference-image sequence of 27 July 2010; paths A and B and the two arrow colours are the source annotations. (b) The corresponding time--distance image along slice A, including the leading track F1, the trailing track S1 that approaches stationarity, and the later track F2. Panels reproduce Figure~1 and the upper panel of Figure~2 of \citet{ChenWu2011}. The F/S labels identify published brightness tracks, not RMO-assigned MHD families.}\label{fig:association}
\end{figure}

E02 takes us back to an earlier question: when an image contains two moving ridges, do we have two local plasma transitions to test? The AIA 193~\AA{} sequence shows a leading ridge and a slower trailing ridge along the same path. The trailing ridge eventually approaches stationarity, and a further branch appears later. Figure~\ref{fig:association} shows this evolution; Appendix~\ref{app:associationdetails} lists the published track speeds.

The ridges separate as they move. They are therefore not simply different sectors of one track or a fixed pattern travelling as a whole. They could belong to a deforming disturbance, a propagating front with changing loops behind it, or several emission ridges within the eruption. The nearly stationary trailing phase must also be distinguished from its earlier motion.

The slower image feature is not automatically a slow magnetosonic wave. In a different event, \citet{Sun2022} interpreted a faster wave train and slower apparent loop expansion within a hybrid EUV-wave model. The slower component there represented the progression of field-line stretching.

For E02, the tracks describe brightness evolution, but do not provide the local three-dimensional normals or the upstream and downstream vector states. We therefore cannot yet construct and test the two states for either ridge. Neither track has an RMO family classification. The next step is to associate each ridge with a plasma response and enough local geometry and state information to test it separately.

\subsection{13 June 1998 (E08) --- Separating the Front and Filament Spectra}\label{sec:e08}
E08 asks whether a moving front and a strong Doppler signal belong to the same structure. The co-registered images show a weak outer front crossing the Coronal Diagnostic Spectrometer (CDS) sampling region and passing beneath the Solar Ultraviolet Measurements of Emitted Radiation (SUMER) slit. No spectral signature of the front is detected. Strong Doppler components appear later, but they belong to the erupting filament.

The timing explains why this distinction matters. SUMER integrates for eight minutes, so a brief front passage can be mixed with emission from other plasma. Appendix~\ref{app:associationdetails} gives the exposure intervals and reported velocities. A small observed line shift need not mean that the plasma associated with the front moves slowly.

For a simple optically thin mixture of stationary background and front emission, the centroid is
\begin{equation}
 v_{\rm centroid}=f_{\rm line}\,v_{\rm front,LOS},\qquad
 0\leq f_{\rm line}\leq1.\label{eq:dilution}
\end{equation}
Here $f_{\rm line}$ is the fraction of line emission supplied by the front. If that fraction is very small, even rapid motion can leave a small centroid shift. Without a positive lower bound on the fraction, the centroid cannot give a finite upper bound on the front-associated plasma velocity. Projection and motion along the front add further uncertainty. This relation illustrates dilution; it is not a fitted correction to E08, and a Gaussian fit to a blended line need not equal this centroid.

The later, stronger velocities and the published line-blend analysis describe the filament. They cannot be combined with the outer-front image speed to form one local plasma state. Nor do they establish a hot component or a temperature jump at the earlier front.

The result is a separation of structures, rather than a shock classification. The spectral non-detection neither selects a family nor shows that no plasma disturbance occurred. A calibrated spectrum isolating the front passage is needed before its upstream and downstream states can be assigned and tested.

\FloatBarrier
\subsection{22 June 2015 (E09) --- A Slow-Shock Benchmark and Its Thermal Model}\label{sec:e09}
E09 starts from a published slow-shock interpretation of dynamical flare loops. The study combined Fe~XXI spectra from the Interface Region Imaging Spectrograph (IRIS; \citealt{DePontieu2014}) with a three-dimensional MHD simulation including heat conduction. We use its processed profiles and a cut through the simulation to ask two separate questions: can we reproduce the spectra, and do the model states have the expected slow-shock ordering?

\begin{figure}[!htbp]
\centering\includegraphics[width=\textwidth]{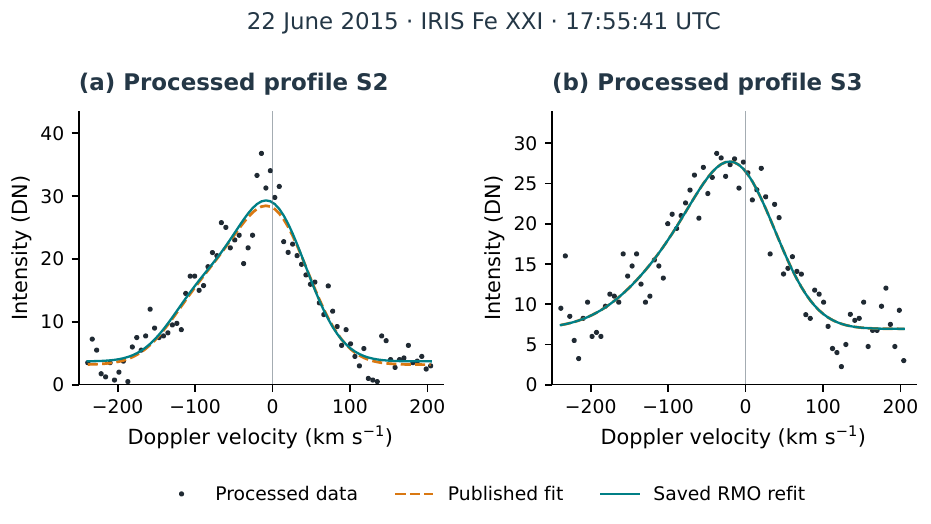}
\caption{E09 spectral reproduction for 22 June 2015 at the source-labelled time 17:55:41 UT. (a,b) Processed IRIS Fe~XXI profiles S2 and S3 supplied by \citet{Ye2026} (points), published fits (orange dashed curves) and retained RMO refits (solid curves). Each curve is a two-Gaussian sum plus background; the refits use the assumed 0.43~\AA{} width floor. All supplied bins are shown. The comparison tests spectral reproduction, not the assignment of the components to upstream and downstream states.}\label{fig:e09}
\end{figure}

Figure~\ref{fig:e09} answers the first question. Independent refits recover the main spectral structure. The S3 decomposition closely matches the published fit; S2 differs slightly, and its narrower component reaches the imposed width limit. The fits remain stable across the tested profile, mask and width choices. Their settings and component centres are given in Appendix~\ref{app:e09details} and Table~\ref{tab:spectra}.

Reproducing a spectrum does not identify which emission comes from plasma before or after a shock. Different decompositions can also remain possible, and undocumented residual weighting cannot be reproduced. We therefore treat this as a spectral reproduction test, not an observed pair of shock states. No new spectral calibration or simulation run is performed.

For the second question, we select plasma samples on the two sides of one interface in the simulation cut. These provide density, pressure and temperature, but only velocity and magnetic-field magnitudes. We use the reported front motion to form conditional shock-frame speeds; the reduced data do not independently establish the necessary signs. Appendix~\ref{app:e09details} records the samples and this transformation.

The thermal model then matters directly. In the isothermal limit, where the characteristic speeds are evaluated using $a^2=p/\rho$, plasma enters faster than the upstream slow speed but slower than the normal \alfven{} speed, and leaves below the downstream slow speed. All nine tested choices of nearby samples show this ordinary slow ordering. They sample one model interface, not nine observed waves.

With adiabatic characteristic speeds instead, the incoming plasma is already below the upstream slow speed, so the same samples do not pass the slow-shock ordering. This does not refute a simulation that includes heat conduction. It shows that the thermal assumptions used in the test must match the physics being examined. Setting $\gamma=1$ in an adiabatic energy equation does not supply the missing conductive heat flux.

The spectra and isothermal-limit ordering therefore support the published interpretation, but do not complete an independent conductive-shock test. The reduced cut lacks the signed vectors and energy fluxes needed for that test, and its selected endpoints also have small mass-flux and normal-field mismatches. Appendix~\ref{app:conduction} explains what must be exported from the model to check the full energy and entropy balance. The next useful step is to recover those local quantities, rather than infer a unique shock classification from the spectra alone.

\subsection{A Compact In-Situ Comparison: Solar Orbiter}\label{sec:insitu}
Solar Orbiter combines remote sensing with in situ measurements of plasma, energetic particles and electromagnetic fields \citep{Muller2020}. Its Radio and Plasma Waves (RPW) instrument also measures local fields and waves and observes solar radio emission \citep{Maksimovic2020}. Relating a remote feature to the plasma sampled by the spacecraft still requires independent association and geometry.

Here we use a switchback boundary to show what direct velocity and magnetic-field measurements add. An ideal rotational discontinuity links their changes through an \alfven{}ic relation. A Wal\'en test compares the measured velocity changes with the magnetic changes expressed in \alfven{}-speed units; its interpretation depends on the reference frame, sampling and plasma assumptions.

For the selected boundary on 30 August 2021, the independent calculation recovers a strong positive relation, close to the published result. Its slope changes with the interval included around the boundary (Appendix~\ref{app:insitudetails}). These interval choices are not error bars. Most importantly, the boundary core contains only one proton sample.

The correlation is reproduced, but the discontinuity type remains unresolved. An independent classification also needs resolved states on both sides, a local normal, and a sufficiently specified reference frame and pressure treatment. Parker Solar Probe also provides in situ plasma and magnetic measurements \citep{Velli2020}, allowing similar tests where both sides of a boundary are adequately sampled.

\section{Which Measurements Distinguish MHD Families?}\label{sec:discriminators}
The examples identify two practical needs: measurements that distinguish the tested shock alternatives, and observations that establish the local states in the first place. We now express these needs in terms of geometry, plasma flow, magnetic-field changes and spectral response.

\subsection{Local Geometry and Upstream Flow}
Reconstructing a front from several viewpoints can constrain its three-dimensional shape and motion \citep{Podladchikova2019,Feng2020}. To distinguish families, we need the local normal relative to both the observing direction and the magnetic field. A projected tangent or a qualitative field plot does not determine these three-dimensional directions and their uncertainties. The required geometrical precision is useful only if the associated flow and thermal assumptions are also justified.

E05 shows why geometry alone may not distinguish families when upstream flow is allowed. E11 gives a quantitative requirement for an independent measurement. The flow test must allow for uncertainty in the front speed, geometry and plasma conditions together. To exclude ordinary slow shocks, the incoming speed must remain above the model's limiting speed for every justified combination of inputs. Comparing only the best-fit values is insufficient. The $277.5\kms$ bound applies to the adopted fixed normal speed and input range; other choices require the joint test in Appendix~\ref{app:e11details}. Failure to establish this sufficient condition does not prove slow or reject fast.

The additional measurement must constrain upstream material motion, rather than another brightness-pattern or radio-front speed. For E11 it must also refer to the patch carrying the radio compression; only then can the flow measurement complete the conditional exclusion test.

\subsection{The Sign of the Magnetic-Strength Jump}
For ordinary fast and slow shocks, the magnetic-field strength changes in opposite directions. This follows from tangential momentum balance, magnetic induction and continuity of the normal field. Appendix~\ref{app:magnetic} gives the relation and its nondegenerate limits.

The observational test is therefore a measured increase or decrease in magnetic strength between the two states. An increase excludes the ordinary slow interpretation; a decrease excludes the ordinary fast interpretation. This test does not establish that a shock is present, classify a restructuring pattern or exclude every other discontinuity family.

If $\widehat{\Delta B}$ is the associated difference in field strengths and $\epsilon_{B1}$ and $\epsilon_{B2}$ are justified bounded absolute errors, a sufficient sign criterion is
\begin{equation}
 \epsilon_{B1}+\epsilon_{B2}<|\widehat{\Delta B}|.\label{eq:precision}
\end{equation}
Calibration, association and shared errors must be included. More generally, the entire supported difference interval must lie on one side of zero. The required precision depends on the field-strength change of the candidate transition. Appendix~\ref{app:e05details} gives planning values for the two E05 examples; they are not measured uncertainties or universal instrumental requirements.

\subsection{Spectroscopy and the Thermal Model}
Spectroscopy is most useful for distinguishing families when it isolates emission from the front and resolves its passage in time. The line response must be separated from reference emission, blends, background and erupting material. E08 illustrates the difficulty of isolating the front response; E09 shows why resolved spectral components still need to be assigned to the two plasma states.

When conduction matters, the energy and entropy fluxes must be evaluated with the model's heat-transport equations. Nearly equal temperatures on the two sides do not determine the conductive flux. Local velocity and magnetic vectors and energy fluxes saved from an existing simulation may then be more useful than additional intensity profiles. The next step depends on which interpretations remain possible under the stated assumptions.

\section{Discussion}\label{sec:discussion}
\subsection{Ambiguity as a Physical Result}
The examples leave different questions open. Several tested connections may satisfy the constraints, or the observations may not yet define the two states. A thermal model may also require energy fluxes absent from the available data. Each situation calls for a different next step: distinguish the surviving connections, establish the two states, or recover the missing fluxes.

The first two examples turn differences between candidate solutions into requirements for flow or magnetic-field measurements. The other examples identify what is needed before a family can be determined. Keeping the assumptions with each result allows the test to be revised when new measurements become available.

\subsection{Relation to Forward Models, Shock Polars and Riemann Solvers}
Forward models and inverse tests can inform each other (Section~\ref{sec:forward}). A model can supply local states, geometry and fluxes at a chosen interface; RMO can identify which quantities need an independent observational constraint. The conductive benchmark shows why a plot of selected variables is insufficient: the test needs the quantities entering the energy balance at that interface.

Shock polars and conventional Riemann solvers can also supply candidate connections for specified states. RMO tests these connections over the allowed input ranges and records the assumptions each requires. In the gas-dynamic fan in Figure~\ref{fig:hd_fan}, a local test concerns the states immediately beside one front. A full solution must also connect them to the rest of the pattern. The local test can check part of a proposed global evolution, while limiting each observational claim to the connection actually tested.

The same distinction applies to structured solar-wind observations. Standing shocks, tearing-driven density structures and oscillatory modulation are all physically plausible in appropriate models \citep{Habbal1985,Reville2020,PodladchikovaPDS2026,PodladchikovaModulation2026}; a recurring brightness pattern does not by itself choose among them.

\subsection{Relation to Detection, Catalogues and Larger Archives}
Automatic detection and tracking establish where a front appears and how it moves. Solar Demon, for example, detects flares, dimmings and EUV waves in SDO/AIA images \citep{KraaikampVerbeeck2015}. JHelioviewer helps users inspect images from several instruments and viewpoints \citep{Muller2017}. Detection and visual comparison help select front patches for local physical tests.

A useful next step is to link each tracked front to plasma diagnostics on its two sides. Where the full states cannot be recovered, a catalogue can retain the observables that constrain them, together with their locations, times, uncertainties and model dependencies. Intensity images alone will not generally supply both MHD states; associated spectra, magnetic information or justified model inputs may be needed. This would connect automatic detection to tests of the physical transition.

Applying the approach to large archives also requires clear selection rules, documented association quality, calibration and uncertainty conventions, and records of null and unresolved cases. These are needed to estimate how often different families occur on the Sun.

\section{Conclusions}\label{sec:conclusion}
Solar fronts are easier to interpret when their motion and brightness are related to the plasma on their two sides. The Riemann picture provides a natural basis for this: a front may mark one transition within a wider evolving pattern. Our contribution is an observational workflow that connects partial measurements to established MHD conditions.

E05 retains both ordinary fast and slow interpretations under the tested assumptions. E11 supplies a flow bound that would exclude ordinary slow shocks if it were independently established and the radio compression belonged to the selected EUV patch. The other examples show where association, thermal assumptions or sampling limit a classification. These results concern local connections; a complete eruption wave pattern requires further information.

Published MHD methods in MPI-AMRVAC \citep{Keppens2023}, PLUTO \citep{Mignone2007} and Athena++ \citep{Stone2020} already use Riemann solvers and have documented benchmark tests. RMO offers a web and Python implementation of the proposed diagnostic workflow, which can also be followed with other suitable solvers or independent calculations. Linking automatic front detection to measurements that constrain both plasma states would help observers and catalogue developers apply this shared body of MHD methods to physical classification and plan the next observation.

\begin{acks}
We thank the researchers who developed the observations, plasma diagnostics, MHD theory and numerical methods on which this work builds. The original observational analyses and simulations discussed here are credited to their authors. We acknowledge the Hinode/EIS, SDO/AIA, GOES/SUVI, STEREO, SOHO, TRACE, IRIS and Solar Orbiter instrument and data teams, and the atomic-data and numerical resources identified in the retained scientific records. The opening SUVI image is credited to Dan Seaton, NCEI and CIRES; SOHO is a project of international cooperation between ESA and NASA.

\end{acks}

\begin{dataavailability}
The observational and simulation source products are identified in the cited publications; the additional Ye et al.\ dataset is identified by \url{https://doi.org/10.6084/m9.figshare.30371113}. The accompanying RMO scientific archive (R139) is available at \url{https://doi.org/10.5281/zenodo.22742107}.
\end{dataavailability}

\begin{codeavailability}
RMO can be used online at \url{https://rmo-solar.org/} or run locally as a Python web application. The source code (v1.0.0-rc3) is archived at \url{https://doi.org/10.5281/zenodo.22741502}. The development repository and installation instructions are available at \url{https://github.com/epodlad/RMO}.
\end{codeavailability}

\appendix
\section{Details of the Physical Tests and Solar Examples}\label{app:checks}
This appendix records the assumptions, numerical inputs and checks behind the main-text results. The event subsections keep each measurement with the choices needed to use it. Numerical tolerances and counts describe the calculations; they do not assign probabilities to the MHD families.

\subsection{Conservation and Characteristic Ordering}
For a locally planar discontinuity moving with normal speed $D_n$, the conservation statement is
\begin{equation}
 [\mathbf F_n(\mathbf U)-D_n\mathbf U]=0,
\end{equation}
where $\mathbf U$ contains the conserved MHD variables, $\mathbf F_n$ is their normal flux and brackets denote downstream minus upstream. In addition, $[B_n]=0$. A consistent frame is used for all components; normal inflow magnitudes used for ordering must not be mixed with laboratory-frame signed fluxes.

Writing $c_{A,n}=|\mathbf B\cdot\mathbf n|/\sqrt{\mu_0\rho}$ and using positive shock-frame crossing speeds, the ordinary fast transition has $w_1>c_{f1}$ and $c_{A,n2}<w_2<c_{f2}$. The ordinary slow transition has $c_{s1}<w_1<c_{A,n1}$ and $0<w_2<c_{s2}$. The retained checks also impose the relevant other inequalities, positive thermodynamic states, compression and entropy increase. These orderings do not replace conservation or a separate treatment of degenerate states.

\subsubsection{Magnetic-Strength Change}\label{app:magnetic}
For a regular compressive ideal-MHD shock, tangential momentum and induction relate the tangential fields on the two sides. With $r=\rho_2/\rho_1>1$ and nonzero normal field, define $A=\mu_0\rho_1w_1^2/B_n^2$. For the nondegenerate coplanar construction, let $B_{t1}\ne0$ and $B_{t2}$ be signed tangential components along a common direction. Then
\begin{equation}
 q\equiv\frac{B_{t2}}{B_{t1}}=\frac{r(A-1)}{A-r}.
 \label{eq:q}
\end{equation}
An ordinary fast shock has $A>r$, hence $q>1$. An ordinary slow shock has $0<A<1$, hence $0<q<1$. Normal-field continuity makes the sign of $|\mathbf B_2|^2-|\mathbf B_1|^2$ the same as that of the tangential-field squared change. The exactly perpendicular and switch limits require their own limiting relations; Equation~(\ref{eq:q}) is not evaluated at its pole or with an undefined denominator.

\subsection{Numerical Safeguards}
E05 uses a complete quadratic construction at each declared model node, recording complex and rejected roots as well as accepted ones. There is no finite magnetic-strength cutoff. The retained limits include $|\mu|\geq10^{-3}$ with $\mu=-\mathbf n\cdot\hat{\mathbf l}_{\rm LOS}$, normalised polynomial residual $10^{-12}$, scaled flux residual $10^{-10}$, characteristic/coincidence margin $10^{-7}$ and relative rational-pole distance $10^{-6}$. The maximum accepted independent scaled flux residual is approximately $3.17\times10^{-12}$. The original unresolved boundary cells remain unresolved; a coarser robustness grid does not certify their interiors.

E11 combines finite constructions with exact rational/Bernstein certificates over stated ordinary-shock domains. Its saved outer extension contains 10,440 nodes and 20,880 algebraic root records; the accepted records include 5,580 fast and 305 slow solutions under separately labelled scenarios. The inner shared-state grid contains 6,660 nodes, with 2,470 accepted slow records and no accepted ordinary fast record in that finite grid. That absence is not an exclusion over untested intervening states. The largest accepted scaled flux residual across the two tests is $3.48\times10^{-15}$. Numerical guard regions and exact limiting states are not forced into a family by the continuous certificates.

\subsection{E05: Inputs, Fits and Measurement Requirements}\label{app:e05details}
\paragraph*{Observations.}
The retained target is EIS exposure 25, central rows 334--338, integrated from 14:28:40.646 to 14:29:25.646 UTC. Reference exposures 1--10 give a Fe~XIII centroid response of $+16.5596\kms$, with a formal noise-only uncertainty of approximately $1.136\kms$. References 1--5 and 6--10 give $+14.5921$ and $+18.5271\kms$, respectively. These alternatives share the target spectrum and are not independent measurements or endpoints of a confidence interval.

Across the 16 retained registrations, the AIA aperture is $2.4$--$8.5\%$ brighter than the early imaging reference. This range describes alignment and exposure choices, not a confidence interval. No unique local ridge, three-dimensional normal or independently associated upstream parcel was established.

The inferred spectral density depends on the assumed temperature. Table~\ref{tab:e05inputs} gives the linked central-aperture values. The published mean speeds of $336\kms$ from AIA and $371\kms$ from EIS are alternative speed proxies, not exact measurements of an instantaneous local normal speed.
\begin{table}[!htbp]

\caption{Linked temperature and density choices for the central E05 aperture. Temperatures are assumed; the densities follow from the corresponding atomic response. These are alternatives, not measured temperature bounds.}\label{tab:e05inputs}
\small
\begin{tabular}{@{}cc@{}}
\toprule
Assumed temperature (MK) & Effective density ($10^8\,\mathrm{cm^{-3}}$)\\
\midrule
$1.0$ & $1.92926$\\
$1.584893$ & $2.35054$\\
$2.5$ & $2.84208$\\
\bottomrule
\end{tabular}
\end{table}

\paragraph*{Sampling.}
The calculation covers 54 linked input sets: three apertures, three references sharing the target spectrum, three temperature/density choices and two speed proxies. A further 18 central-aperture sets vary the assigned upstream normal flow. Both ordinary families have accepted examples in every one of these 72 sets.

The 76,842 deterministic model cases include 703 reused cases from the unchanged baseline. They generated 112,644 algebraic root records and 36,937 accepted records: 33,933 \fast{} and 3,004 \slow{}. These are counts of model realisations, not probabilities or fractions of parameter-space volume.

\paragraph*{Reference-spectrum fits.}
The two examples use the same assignment of plasma samples to shock states and are compared with ten reference spectra. Their assumed $(\alpha,\theta)$ angles are $(14^\circ,15^\circ)$ for \fast{} and $(73.875^\circ,51^\circ)$ for \slow{}. Here $\alpha$ is the front-normal tilt in $V_n=V_{\rm proxy}\cos\alpha$, and $\theta$ is the field-to-normal angle. Neither is measured.

The assumed tangential velocity satisfies $\mathbf v_t\parallel\mathbf n\times\hat{\mathbf l}_{\rm LOS}$, where $\hat{\mathbf l}_{\rm LOS}$ is the line-of-sight unit vector. Its Doppler contribution, $\mathbf v_t\cdot\hat{\mathbf l}_{\rm LOS}=0$, was checked at every accepted state. This follows from the assumed geometry; it is not an independent measurement of zero tangential velocity.

The fast and slow fits have excess chi-squared values of approximately $0.01720$ and $0.01574$ relative to their respective conditional free-ratio models. The comparison neither gives a global likelihood ranking nor proves that the reference is upstream.

\paragraph*{Geometry and magnetic measurements.}
With stationary upstream plasma, the boundary near $\alpha=54.227^\circ$ gives sufficient planning ranges: $45^\circ\pm5^\circ$ excludes the checked slow family, and $60^\circ\pm5^\circ$ excludes the checked fast family. Allowing upstream motion gives a \slow{} example at $45^\circ$ with assigned flow $+53.512\kms$, and a \fast{} example at $60^\circ$ with $-44.207\kms$. Neither the angles nor the flows are measured; tilt alone no longer distinguishes the families.

The saved examples in Figure~\ref{fig:e05}c have $(B_1,B_2)=(0.925217,1.023508)\gauss$ for \fast{} and $(3.733122,3.715630)\gauss$ for \slow{}. No independent field measurements on both sides of the same patch are available for this test.

The two E05 examples would require equal per-state absolute errors below approximately $0.049146\gauss$ and $0.008746\gauss$, respectively. These planning values apply only to the two conditional examples. They are not measured uncertainties, universal instrumental requirements or lower bounds over all weak and degenerate limits.

\subsection{E11: Local Inputs and the Conditional Flow Bound}\label{app:e11details}
\paragraph*{Geometry and plasma inputs.}
The selected outer AIA 193~\AA{} crest is near position angle $116^\circ$. Its fitting interval is 05:38:32.06--05:40:08.06 UTC, and the adopted geometry gives a fixed normal pattern speed of $705.535\kms$.

For the E11 G0 geometry, the retained unit normal is
\begin{equation}
 \mathbf n=(0.8636020941,-0.5041740007,0),
\end{equation}
in axes toward solar west, solar north and the observer. Its zero line-of-sight component is why a Doppler velocity alone cannot supply the upstream normal flow used in the main-text closure test.

The adopted inputs are lower and upper radio frequencies $f_L=127$--$137\,\mathrm{MHz}$ and $f_U=150$--$180\,\mathrm{MHz}$, with $T_1=1.4$--$2.2\mk$. Under the adopted emission model, assigning the two bands to upstream and downstream plasma gives the density compression $r=\rho_2/\rho_1=(f_U/f_L)^2$. For the central inputs, $r=1.5625$, $\rho_1=9.0351\times10^{-14}\,\mathrm{kg\,m^{-3}}$ and $T_1=1.8\mk$. We calculate pressure for fully ionised hydrogen with equal electron and proton temperatures and $\gamma=5/3$. These inputs retain their shared dependencies; their ranges do not specify a joint observational confidence region.

The ordinary fast branch is established for $0<\theta_{Bn}\leq90^\circ$; the exactly parallel switch limit remains degenerate. A published potential-field source-surface (PFSS) plot suggests a nearly perpendicular geometry but supplies neither a numerical local magnetic vector nor a bound on the field-to-normal angle at the selected patch.

\paragraph*{The flow threshold.}
The limiting incoming speed in this model is
\begin{equation}
 w_{\rm crit}=\sqrt{\frac{5(p_1/\rho_1)r}{4-r}}.\label{eq:wcrit}
\end{equation}
For the central inputs, $w_{\rm crit}=308.615\kms$ and the upstream-flow threshold is $u_{\rm crit}=396.921\kms$ at the adopted fixed normal speed. The main-text bound of $277.5\kms$ is a slightly conservative rounding of the saved $277.513560\kms$ boundary across the adopted radio and temperature domain. No such bound has been measured, and applying it requires the radio compression to belong to the selected EUV patch.

Below the relevant upstream-flow threshold, ordinary fast solutions exist and ordinary slow solutions are excluded. With no independent bounds on angle or field, slow solutions can be constructed arbitrarily close above onset. A central-input \slow{} example at $\theta_{Bn}=23^\circ$ has assigned $u_{1n}=427.782\kms$, $w_1=277.753\kms$ and $w_2=177.762\kms$. It preserves the source compression and temperature and passes entropy and flux checks; the assigned flow is not an observation.

If normal speed, upstream flow and thermodynamic inputs vary together, a sufficient slow-exclusion condition is
\begin{equation}
 \inf_{\rm supported}\left[(V_n-u_{1n})-w_{\rm crit}\right]>0.
 \label{eq:joint}
\end{equation}
The infimum is the greatest lower bound over the observationally justified joint input combinations. The margin must remain positive throughout that domain. Changing the normal or allowed front speeds requires reevaluating this condition. Failure to establish it does not prove slow or reject fast.

\paragraph*{Other connections and the inner ridge.}
The E11 field-reversing candidates fail the unrestricted ideal-MHD transverse evolutionary test. Two independent out-of-plane jump conditions must be satisfied, but the characteristic count provides only one outgoing \alfven{} amplitude for three incoming amplitudes. Generic incoming perturbations therefore cannot be matched. This result applies to the examined coplanar base states with unrestricted perturbations; it does not address dissipative stabilisation or perturbations restricted to the plane. The contact and ideal rotational exclusions in Section~\ref{sec:e11} follow from the imposed compression and finite mass flux. These model exclusions do not rule out a material boundary as an image interpretation when the radio source has not been associated with the patch.

Over the same fitting interval, the inner and outer ridges have projected speeds of $409.1\kms$ and $687.5\kms$, respectively, distinct from the adopted $705.535\kms$ normal speed. The saved inner \slow{} example assigns the outer model's downstream state to the plasma between the ridges. This uniform state is not measured, so the two tracks do not establish a solar fast--slow pair or a complete Riemann fan.

\subsection{E02 and E08: Track and Exposure Details}\label{app:associationdetails}
\paragraph*{E02 image tracks.}
The published AIA 193~\AA{} slice A contains an initially faster track near $560\kms$ and a slower track near $190\kms$ that decelerates towards stationarity. Slice B contains tracks near $470$ and $170\kms$. A later branch in slice A is reported near $310\kms$. F1--F3 and S1--S2 are source labels for image tracks, not RMO mode assignments. The great-circle paths and PFSS context do not provide local three-dimensional normals or upstream/downstream vector states for either ridge.

\paragraph*{E08 spectral sampling.}
The reported SUMER mode used $480\,\mathrm{s}$ integrations. The first approximate interval, 15:31--15:39 UT, includes the front passage without a detected wave signature. The following 15:39--15:47 UT interval contains blue-shifted transition-region filament emission. The retained CDS account places the weak front's line-of-sight response below approximately $10\kms$. This response scale depends on the instrument and sampling; it is not a bound on intrinsic normal flow or a new confidence interval.

The later components near $150$ and $300$--$350\kms$ belong to the filament. The published Mg~X 609.79~\AA/O~IV 609.83~\AA{} blend analysis also concerns filament emission. It supplies neither a hot component nor a temperature jump for the earlier outer front. The dilution relation in Equation~(\ref{eq:dilution}) is an illustration, not a fitted correction to these data.

\subsection{E09: Spectral Fits and Model Samples}\label{app:e09details}
\paragraph*{Spectral reproduction.}
The refits use a constant background and an assumed minimum full width at half maximum (FWHM) of 0.43~\AA. Table~\ref{tab:spectra} lists the component centres. The width floor and undocumented residual weighting limit the comparison described in Section~\ref{sec:e09}.
\begin{table}[!htbp]
\caption{E09 Doppler centres copied from the saved reproduction ($\mathrm{km\,s^{-1}}$). Fit outputs are shown without measurement error bars. The refit uses a constant background and an assumed 0.43~\AA{} width floor.}\label{tab:spectra}
\small
\begin{tabular}{lll}
\toprule
Profile & Published curves recovered & Independent refit\\
\midrule
S1 & approximately $+8.0$ & $+8.012$\\
S2 & $-92.834$, $-2.275$ & $-94.984$, $-4.018$\\
S3 & $-78.503$, $-7.206$ & $-78.503$, $-7.207$\\
\bottomrule
\end{tabular}
\end{table}

\paragraph*{States and thermal limits.}
The 320-point model cut is at $1.06t_0$, distinct from the synthetic spectral samples at $1.08t_0$. Here $t_0$ and $L_0$ are the source time and length scales. The cut contains density, pressure, temperature, an absolute normal velocity and magnetic magnitudes. The representative endpoints are listed in Table~\ref{tab:e09model}; their density, pressure and temperature ratios are $1.3558$, $1.3957$ and $1.0295$.

The conditional shock-frame speeds use the reported approximately $258\kms$ front motion opposite to positive evaporation. Magnitude columns alone cannot determine this signed transformation.
\begin{table}[!htbp]

\caption{Representative E09 model samples. Speeds are in $\mathrm{km\,s^{-1}}$. The shock-frame transformation depends on the reported front motion and assumed signs. The two choices of characteristic speed test different thermal limits.}\label{tab:e09model}
\small
\begin{tabular}{@{}lrr@{}}
\toprule
Quantity & Upstream & Downstream\\
\midrule
Position ($L_0$) & $0.018$ & $0.038$\\
Shock-frame speed magnitude & $456.18$ & $342.84$\\
Slow speed, isothermal limit & $366.79$ & $381.34$\\
Slow speed, adiabatic limit & $472.30$ & $490.56$\\
\bottomrule
\end{tabular}
\end{table}

For the isothermal-limit check, Equation~(\ref{eq:chars}) uses $a^2=p/\rho$; the adiabatic check uses $a^2=(5/3)p/\rho$. All nine nearby endpoint pairs across this one interface pass ordinary slow ordering in the isothermal limit, and none passes it in the adiabatic test with the adopted shock-frame transformation. The contrast concerns the diagnostic thermal assumptions, not nine observed waves or a rejection of the conductive simulation.

The selected endpoints differ in mass flux by approximately $1.89\%$ and in normal-field magnitude by $4.24\%$. The reduced cut does not show how much of these differences comes from background variation, time dependence, transverse geometry or a local jump. A nonuniform background does not remove the normal-field continuity condition for a properly defined discontinuity.

The heat-conduction analysis recovered the published thermal model. The missing signed vectors and energy fluxes needed for an independent full test are specified in Appendix~\ref{app:conduction}.

\subsection{Solar Orbiter: Boundary Sampling and Wal\'en Slopes}\label{app:insitudetails}
For the saved Event 2 CS2 boundary on 30 August 2021, 10:21:24--10:21:28 UT, the published slope is $+0.973$. The retained independent calculation from the same-day data files gives $+0.9248$ with a 15-s context extension. The 10-, 15- and 20-s alternatives give approximately $+0.776$, $+0.925$ and $+0.974$. These are interval choices, not error bars. The 15-s context contains nine proton samples, but the boundary core itself contains only one.

\section{Information Needed for the E09 Conductive Energy Balance}\label{app:conduction}
The nearly equal endpoint temperatures do not determine the conductive heat flux. This appendix identifies the terms needed to test the model's energy and entropy balance independently.

\subsection{Ideal and Conductive Fluxes}
The working analysis retains the source model's $\gamma=5/3$ gas and anisotropic thermal conduction. In dimensional cgs units, its ideal normal energy flux contains
\begin{equation}
\begin{split}
 F_{{\rm ideal},n}={}&\left[\frac{\gamma P}{\gamma-1}
 +\frac{\rho(V_n^2+|\mathbf V_t|^2)}{2}
 +\frac{|\mathbf B_t|^2}{4\pi}\right]V_n\\
 &-\frac{B_n}{4\pi}\,\mathbf V_t\cdot\mathbf B_t.
\end{split}\label{eq:idealflux}
\end{equation}
Here $P$ is gas pressure and $\mathbf V$ is the signed laboratory plasma velocity in the model; the notation is local to this appendix. The reduced scalar columns do not fix $|\mathbf V_t|^2$ or $\mathbf V_t\cdot\mathbf B_t$. Even the ideal contribution is therefore undetermined before conduction is added.

For a field unit vector $\mathbf b$, the unsaturated conductive normal flux is
\begin{equation}
 q_n=-\kappa_\perp\partial_nT
 -(\kappa_\parallel-\kappa_\perp)b_n
 \left(b_n\partial_nT+\mathbf b_t\cdot\nabla_tT\right).
 \label{eq:heat}
\end{equation}
A one-dimensional temperature trace does not determine the transverse gradient in this expression. The original saturation rule and normalisation must also be retained. A scalar saturation scale is not evidence that a face attains that flux.

\subsection{Energy and Entropy Balance}
For a moving control volume with boundary velocity $\mathbf v_b$, a laboratory-frame energy-balance check would use
\begin{equation}
 \frac{\mathrm d}{\mathrm dt}\int_{\Omega(t)}e\,\mathrm dV+
 \oint_{\partial\Omega(t)}(\mathbf F_E-e\mathbf v_b)\cdot\mathbf n_{\rm out}\,\mathrm dA
 =\int_{\Omega(t)}\rho\mathbf g\cdot\mathbf V\,\mathrm dV,
 \label{eq:control}
\end{equation}
where $e$ is the total energy density excluding gravitational potential energy, $\mathbf g$ is gravity, and $\mathbf F_E$ includes ideal, conductive and resistive fluxes. Storage, lateral fluxes, geometry and gravitational work can be dropped only with independent justification. The entropy balance additionally includes the transported heat-entropy flux $\mathbf q/T$. The saved gas-entropy proxy of approximately $-0.174$ between the representative endpoints is consequently not, by itself, a rejection of a conductive shock.

\subsection{Information to Export from the Model}
A sufficient export would contain the oriented patch and sampling faces near the same $1.06t_0$ cut; signed vector states and consistent units; conductive and resistive face fluxes or the stencils and implementation needed to recover them; and adjacent-time and lateral terms, or a justified steady planar reduction. No numerical energy or entropy residual was calculated in the working analysis because these inputs were unavailable. Source notation and unit inconsistencies recorded there require confirmation before a future flux reconstruction; they are not evidence of a simulation error.

\subsection{Additional Spectral Checks}
These checks concern profile reproduction, separately from the conductive balance. Subtracting the supplied blend curves from the original profile columns reproduces the cleaned S2 and S3 profiles to better than $1.1\times10^{-14}$ detector counts (DN). Two ancillary synthetic P2 component curves miss the preset $10^{-5}$ reduced-reconstruction tolerance by $0.068\%$ and $0.415\%$. These auxiliary checks are not used to reject the published slow-shock interpretation.

\bibliographystyle{spr-mp-sola}
\bibliography{references}
\end{document}